# Quantum Effects and Genetic Code: Dynamics and Information Transfer in DNA Replication

S. Mayburov

*Lebedev Inst. of Physics*

*Leninski pr. 53, Moscow, Russia, 117924*

*E-mail: mayburov@sci.lebedev.ru*

C. Nicolini , V. Sivozhelezov

*Nanoworld Inst. and Biophysics Chair, University of Genoa*

*Corso Europa 30, 16132 Genoa, Italia*

## Abstract

The possible role of some quantum-mechanical effects in the transfer of genetic information during the cells division is studied. It's argued that the efficient nucleotides selection in DNA replication is performed via the protons tunnelling between the nucleotide and DNA-polymerase which bind them by hydrogen bonds. Such mechanism is sensitive to the structure of nucleotide hydrogen bonds, consequently only the particular nucleotide sort will be captured by DNA-polymerase, depending on its state. This multistep selection mechanism is also analysed as the algorithm of data base search, where the data base is the pool of similar molecules. It's shown that for simple models it's analogous to Grover algorithm, which can be realised both in classical and quantum dynamical systems.

## 1. Introduction

DNA replication and proteins synthesis are the vital processes that form the basis of life and evolution. DNA replication and repair provide the transfer of genetic information through the long sequence of cell divisions with high accuracy. Simultaneously it affords the minor controlled rate of mutations which stipulates the permanent evolution and diversity of species. The mechanism of replication is quite complicated and despite the great achievements of last sixty years, many of its features are still poorly understood (Lodish et al., 2000). In this paper we argue that some quantum-mechanical (QM) effects, in particular the proton tunnelling, play the important role in the correct nucleotides selection during DNA replication. The algorithm of this multistep process will be regarded also in the framework of Computation Theory from the point of its efficiency, such analysis is especially useful for the comparison of quantum and classical selection mechanisms.

Remind that DNA molecule basically consists of two parallel molecular strands, each of them contains 1-dimensional chain of nucleotide fragments or bases – Adenine (A), Thymine (T),Cytosine (C) or Guanine (G) connected via the phosphate and sugar groups (Watson, 1987). Each such sequence of nucleobases encodes the necessary genetic information, other strand just doubles it as the mirror copy. This information is recorded along the strand, alike the bits on the magnetic tape, each base is roughly equivalent to two bits. During the protein synthesis the encoded messages are transferred to the ribosome which transforms it into the receipts of protein assembling. DNA strands are bound via all its nucleobases, but only A-T and C-G pairs constitute the chemical hydrogen bonds between them, it results in the mentioned information equivalence of both strands. During its replication DNA base pairs are unzipped into this strands, and each strand is processed by one of DNA - polymerase (DNA-pol) enzymes. The acquired information is used to assemble the second strand from the specific moleculas floating in the intercellular media; eventually new DNA molecule is constructed. Such floating molecules - deoxynucleotide triphosphates (dNTP) includes one of mentioned A,T,C,G nucleotides and phosphate group which separated from them during DNA assembling (fig. 1).

Obviously to prepare the exact copy of initial DNA, dNTPs should be identified correctly before to be implanted as the partner of given base template in a DNA strand. The mechanism of such selection is still quite poorly understood, it illustrated by the difficulties in the calculations of fidelity $F$ of DNA replication, i.e. the percentage of errors in a newly constructed DNA molecule. Chemical Kinetics computations predict the value of $F$ about $10^{-1}$-$10^{-2}$, whereas the experiments with the main families of replicative DNA-pols gives $F$ up to $10^{-6} \div 10^{-7}$; the additional improvement by a factor 10 is achieved by a secondary check by exonuclease (Beard, 2002). The experimental data shows that dNTP selection induces the complex sequence of chemical and mechanical processes (Kunkel, 2004). All DNA-pols have the intricated geometric structure, usually their form is analogous to human right hand: the palm, thumb and fingers domain, the fingers domain most closely interact with incoming dNTP. The experiments evidence that during DNA replication DNA-pol passes through the complex sequence of conformations (Kunkel, 2004). Normally the palm is open, but when the base-pairing process starts, the palm closes, resulting in the tight contact between this two nucleotides – so called 'induced-fit' (Sawaya, 1997). Obviously if no preliminary selection for incoming dNTP is done, then after the palm closure more than 80% of these synthesis attempts will not result in the correct pair production, which means large loss of efficiency together with the gain of errors number. Hence if some DNA-pol mechanism can select the incoming dNTPs, before DNA-pol mechanical transformations would begin, it will rise the efficiency of DNA replication considerably. The fidelity at this initial stage can be not too high, it's enough to have $F$ of the order $10^{-2}$. The bulk of experimental data now confirms the existence of dNTP preselection for different DNA-pol types and its close connection with the 'induced-fit' transformation, however its mechanism is still quite obscure (Beard, 2002).

In this paper the possible role of Quantum- Mechanical (QM) effects in dNTP preselection is investigated. The quantum properties of molecules and atoms are widely used in Biology, but usually they are used mainly phenomenologically. We shall argue here that the study of DNA replication demands the more detailed use of QM formalism, in particular, the account of quantum dynamics is necessary. One of arguments in favour of this approach is the surprising resemblance between the parameters of genetic code and the quantum algorithm of data base search (Patel, 2001). From the computational point of view dNTP selection by DNA-pol is the search of data base which in this case is the pool of different molecules in intracellular media. Below we shall describe the model in which biochemical molecules are regarded as the ensemble of oscillators for which the algorithm of data base search is realised by means of specific analogue computations.

Here we shall argue also that dNTP preselection is stipulated by the different structure of hydrogen bonds vacancies in dNTP species and is performed via the proton tunnelling between dNTP and DNA-pol. The quantum tunnelling i.e. the passage of quantum particles under the potential barrier is the direct consequence of QM uncertainty principle. The nucleobase pairs in DNA are connected by 2 or 3 hydrogen (H) bonds, because of it, the proton tunneling between the nearby p-donors or p-acceptor is the potential source of DNA mutations (Lowdin, 1963). The recent experiments demonstrated the crucial role of proton tunnelling in biological catalysis, namely that the excited states of enzymes can effectively reduce the proton potential barriers between reactants (Kohen, 1999). It results in the fast delocalization of protons initially belonged to one of reactants and consequently the rate of reactions with the hydrogen transfer is much higher than standard Theory of Catalysis predicts (Masgrau, 2006). In our case the different dNTPs posses different structure of p-donors and p-acceptors, hence the rates of proton tunneling between them and DNA-pol can be different. Consequently it results in the quite different binding of DNA-pol and dNTP types. Such model of dNTP identification, in which the proton tunneling is regulated by the excited states of DNA-pol enzyme, will be considered below.

## 2. Methods and Model
### A. Data Base Search and DNA replication

It's well known that DNA and RNA performance during the transfer of genetic information can be described by some computation algorithms (Lodish, 2000). In particular, the selection of particular dNTP from a pool of similar molecules is described by the algorithm of data base search (Patel, 2001). In the simplest case such search is the selection of one particular item characterised by the specific value of its discrete parameter $J^p = J_0$ from the array of $N$ analogous items with unknown $J^p$ value. In classical case, to find the sought object, one should perform a one by one check of the array items. It demands on the average to perform the sequence of approximately $N$ operations (queries) on this array with *yes/no* outcomes. Remarkably, the quantum search algorithm which operates with the pure quantum satets can fulfil such search performing about $\sqrt{N}$ queries only (Grover, 1997). The parameters of this algorithm, as was shown recently, astonishingly coincide with parameters of genetic code, in particular single query can check 4 items (objects), which is the number of fundamental DNA nucleotides. In addition 3 queries can check 20 objects, which coincides with the number of basic aminoacids used in the protein synthesis; there are also some other more subtle indications (Patel, 2001).

It was proved that the quantum search algorithm is ultimately optimal for such problems, the similar efficiency is unfeasible for the classical discrete computations. Such optimality makes this algorithm especially interesting in Biology, because it can provide, in principle, the best conditions for the organism survival. For realisation of quantum computations in biosystems the main obstacle is the decoherence of pure quantum states by biological environment, and it seems to be essentially unavoidable (Guilini). Yet it was shown that this search algorithm can be realised also in classical continuous systems, for example in the array of coupled harmonic oscillators. In particular, the time $t$ for the search of array with $N$ objects grows proportionally only to $\sqrt{N}$, and not $N$ as in classical discrete case (Lloyd, 2000). For such systems the decoherence is absent and so it makes them the real candidate for data base search in biosystems. In addition due to the complexity of bio-systems in some cases their effective dynamics is nonlinear, despite that it originates from QM which is linear theory. The example of such nonlinearity is the observation of soliton effects in Bioenergetics (Davydov, 1991); in principle for the specially adjusted parameters the analogous nonlinearity can improve further the efficiency of data base search in comparison with linear system.

Here we shall investigate only the linear systems with a few degrees of freedom, because of their relative simplicity. In this vein Langrangian model for data base search was proposed, the model system, used also here, is the array of coupled harmonic oscillators (Grover, 2002). Note that the similar systems often used also for simulation of molecular dynamics of vibrational and electronic excitations (Davidov, 1991). Consider an array $\Gamma_O$ of $N_O$ pendulums $O_i$ and suppose that all of them, except one with index $j$, has the same length $l_i=l$, the problem is to define $O_j$ with the different length $l_j=l^d$ ( for the simplicity of notations let's fix $j=1$ beforehand). The Langrangian of each pendulum $O_i$ is:

$$L_i = \frac{1}{2}(m\dot{x}_i^2 - k_i x_i^2) \qquad (2)$$

where $k_i = \frac{mg}{l_i}$; $x_i, \dot{x}_i$ are its coordinate and velocity. The classical search - i.e. one by one check of pendulums will take the time $t$ proportional to $N_O$, however the system of coupled oscillators permit much faster check (Grover & Sengupta, 2002). Assume that each $O_i$ is coupled to the large pendulum B which dump its initial energy to the system $\Gamma_O$ of all $O_i$. The Langrangian of B,$\Gamma_O$ system is:

$$L_{BO} = \frac{1}{2}(M\dot{X}^2 - KX^2) + \sum L_i + \sum L_i^B \qquad (3)$$

where $L_i^B$ describes B, $O_i$ interaction; B,$\Gamma_O$ parameters can be varied to become optimal for our aims. Initially at $t=0$ all $O$ are at rest ($x_i=0$, $\dot{x}_i = 0$) ; yet B has large potential energy $U_B$, which corresponds to initial $X=1$. From the simple calculations, at time $t= \tau\sqrt{N}$, where $\tau$ is $O_1$ period, the amplitude $A_1(t)$ of $O_1$ oscillations will be of the order 1, whereas all other $O$ amplitudes $A_i$ at that time will be of the order $\frac{1}{N}$. Hence in arbitrary case the high value of $O_j$ amplitude and its kinetic energy identifies the initially unknown pendulum $O_j$ for time $t$ with high fidelity. The detailed analysis shows that in this case the optimal B,$\Gamma_O$ parameters correspond to the resonance in B,$\Gamma_O$ system which results in the transfer of energy from B to $O_1$ only. Eventually in this resonance regime such system performs the optimal algorithm of data base search with $N_O$ objects and defines unknown object $j$ for the time $t \approx \tau\sqrt{N}$, the obtained search speed is equivalent to one of quantum algorithm. Note that no classical discrete search algorithm can be performed with such efficiency.

For DNA-pol the possible realization of this algorithm can be as follows: suppose that each of mentioned 4 DNA-pol fingers interacts strongly with the incoming dNTP and constitute 4 independent oscillators $O_i$. If each finger parameters are tuned to the resonance with the particular nucleotide C,G,A,T, then it's possible that the single oscillator parameters will be specific to the given incoming nucleotide. Hence if DNA-pol acts as large oscillator B and dump the energy inside this array of fingers, then only one of them will be excited in each event, and the number of this finger $J_f$ defines which dNTP enters DNA-pol volume. Such mechanism will be also quite error-safe: if the excitation threshold is high enough, then the number of incorrectly defined dNTP will be small, so it permits in principle to reach the high fidelity of DNA replication. In particular, such high energy of given $O$ permits to perform the efficient passage of protons over the potential barrier which divide dNTP and DNA-pol and speed ip their bonding. Regarded model is quite simple and abstract, but below we shall see that the analogous features can be found in more complicated and practical model.

## B. Quantum Tunneling in DNA Replication

Now we consider in detail the model of dNTP selection which exploits the quantum tunnelling of protons as the main dynamical mechanism, below this mutistep quantum process will be analysed also as the computation algorithm from the point of its effiency. To illustrate the tunneling effect, consider the free particle of mass $m$ with kinetic energy $E$ directed to the rectangular potential barrier of height $U$ and width $z$. If $E > U$, both the classical and quantum particle would pass over it effectively, yet in quantum case even if $E < U$, some part of particles flow $T$ will penetrate through the barrier and registered on its opposite side. $T$ is called the tunnelling (transmission) coefficient, for the barrier of such form the quantum-mechanical calculations give, neglecting small terms:

$$T(z) \cong e^{-\frac{2z}{\hbar}\sqrt{2m(U-E)}} \qquad (1)$$

where $\hbar$ is Planck constant. The possible importance of proton tunneling for DNA functioning, due to the hydrogen bonding between its strands, was understood in the sixties (Lowdin,1963). Since then the main efforts were concentrated on the study of its role in DNA mutations, which can occur via p-tunneling inside the same nucleotide or between two nucleotides – tautomerisation (Li,2001; Dabrowska, 2004).

Nucleotides in DNA are connected by the hydrogen (H) bonds; such bonds are produced between the atomic p-donors and p-acceptors which denoted here both as pDA . Usually p-acceptor atoms posses the lone pairs of electrons on their external shell; this is typical for electronegative atoms like N,O,F, etc. The p-donors are also mainly electronegative atoms which constitute the stable but relatively weak chemical bond with H. If such molecule and p-acceptor approach closely to each other, the proton can penetrate through the potential barrier between them to the location of p-acceptor atom. In standard conditions, if this barrier is large, i.e. $T<<1$, the rate $R_H$ of H-bond formation is presumably proportional to $T$ of (1) (Joesten,1974). Really to start H-bond formation, the proton should tunnel through the potential barrier to p-acceptor and its probability is proportional to $T$. After the proton depart the p-donor charge is equal to $e^-$, and its electrostatic field will attract back the proton, it effectively corresponds to the initial barrier reduction (see also below). When H-bond is formed finally, the proton oscillates between the donor and acceptor regions; the probability $P_A$ to find it in acceptor region is usually much less than .5. As the result the donor and acceptor attract one another mainly electrostatically, the exchange interactions produce the minor repulsion effect, other terms are small (Joesten,1974).

The majority of biochemical reactions between the reactant molecules $Re_i$ practically can be realised only in the presence of enzymes $En$, the typical reaction scheme in standard Theory of Catalysis i.e. Transition State Theory is :

$Re_1 + Re_2 + En \rightarrow X_R \rightarrow En + Pr_1 + Pr_2 + ...$

where $X_R$ is the transition state, $Pr_i$ are the reaction products. As was mentioned in chap.1, it was found recently that some enzymes can reduce the width of proton potential barrier between reactants; it results in much higher rate for the hydrogen transfer reactions than standard Theory predicts (Masgrau,2006). The physical mechanisms of this effect are still disputed, but there are the strong indication that the p-barrier reduction produced mainly by the excited states of enzymes, in particular, as the result of the internal protein motion (Hammes-Schiffer, 2006). The reactions with H-bonds formation are very similar to the proton transfer reactions, hence the same p-tunneling enhancement can be important for them also, in particular for H-bonding between nucleotides in DNA.

The huge diversity of DNA-pols was found experimentally, they constitute several large families denoted A,B,α,β, etc.. The different families have the fidelity differing by many orders; in our model only dNTP selection for most effective families A,B will be studied (Beard, 2002). As was mentioned above, dNTP selection accompanied by the complicated sequence of DNA-pol conformations (Arora, 2004). However until now the experimental data on them for main DNA-pol families are still incomplete (Kinkel, 2004). In our model the possible scenario of DNA-pol domains motion is chosen so, that it permits to analyse the role of p-tunneling controlled by DNA-pol in a most simple and straightforward way, simultaneously it doesn't contradict to the existing data. However it can't excluded that the real process of selection develops in more complicated and entangled fashion, but even in that case we expect that proposed here selection mechanism can be involved in it, may be in a modified form.

In terms of Catalysis Theory DNA-pol in our model plays in fact the roles both of enzyme $En$ and one of reactants $Re_1$, the other one $Re_2$ is an incoming dNTP. DNA-pol fragments which act as the reactants in this case are supposedly located on its surface as 'hot spots' (HS) of molecular size, whereas the enzyme is the rest of DNA-pol. Each hot spot contains the particular configuration of pDAs $HS_j$ ($j=1,4$) which in principle can bind strongly by H-bonds only dNTP $D_i$, ($i=1,4$) of one particular kind A,T.C or G only ( 'key and lock' relation), hence altogether there should be the hot spots of four different configurations $HS_j$ denoted $A^n, T^n, C^n, G^n$. The total number of different hot spots on DNA-pol surface is approximately the same. $HS_j$ structure will be described below in detail, but in general their pDAs configuration is supposedly analogous to the one of their template partner in DNA; for example to attract T nucleotide, $T^n$ should include at least one p-donor and p-acceptor at the distance about 1.8 A (Angstrom) between them, i.e. to be analogous to pDA structure of A nucleotide (Fig.1). Such hot spots, effectively binding some molecules on the surface of large protein molecules were found in the experiments with protein-protein complexes (Kortemme,2002; Burgoyne,2006). Possible existence of hot spots with some specific functions on DNA-pol surface was already discussed (Beard, 2002).

Concerning with the dynamics of dNTP binding by DNA-pol, we concede that before DNA replication starts all hot spots are in the ground state, such that the potential barriers of their pDAs has a width and height, typical for the biochemical molecules in a ground state, so that any free dNTP practically would not be bound by any of them. DNA-pol at that period is also in the ground state denoted $G_0$. At each replication step DNA-pol unzips one DNA base pair and supposedly identifies the nucleobase $D^B_i$ for which the partner dNTP $D_j$ should be found. Here we assume that depending of this base sort - A,T,C or G, the different DNA-pol state $G_j, j=1,...,4$ will be excited for some period $\tau$. This state can spread in any part of DNA-pol and activate only the set of corresponding hot spots $HS_j$. In this excited spot state the width of $HS_j$ potential barriers for its pDAs are reduced and the protons would pass through them with high efficiency. As was noticed above, the proton barrier reduction is induced namely by an excited thermal states of enzymes, and not by their ground states (Benkovic, 2003). One can expect that the barrier squeezing by the ordered excited states $G$ of enzymes will be even more efficient. Hence if the suitable free dNTP $D_i$ would approach $HS_j$ at that excitation period $\tau$, then H-bonds will be formed promptly between them, and this dNTP will be confined by DNA-pol with high probability. Consequently on DNA-pol surface mainly one of four possible kinds of molecular complexes will appear: T,T$^n$; A,A$^n$; C,C$^n$; G,G$^n$, i.e. in each event the nucleotide which is the correct base partner will be confined with high fidelity. After the excitation period $\tau$ is finished any $G_j$ state supposedly changes to final DNA-pol state $G_{fin}$ which results in DNA-pol palm closing. If the given dNTP will be dispatched in this way to the DNA base strand, then the attempt of DNA base pair synthesis would start. The collective excited states normally are nonlocalized and spread over a large part of the system volume. As was noticed the incoming dNTP contacts first with the fingers domain of DNA-pol, the data shows that it distinctly divided into several subdomains called 'fingers' (Beard, 2002). In our model for the simplicity we suppose that the effective number of this fingers is 4, and the hot spots of the same kind $HS_j$ are concentrated only on the unique finger $j$, so that the state $G_j$ excites only this finger $j$.

To estimate the effect of quantum tunneling for dNTP binding to DNA-pol, BB model can be used; it's widely applied for the calculations of proton tunneling stipulated by enzymes and its formalism is most suitable for our problem (Bruno, 1992). The more elaborated models were proposed later, but the numerical computations for them are more complicated (Benkovic, 2003). BB model assumes that the collective degrees of freedom (DF) for the given enzyme-reactants system can be effectively described by the classical harmonic oscillators. In particular, the width of potential barrier $z$ is controlled by the oscillator $O_E$ of mass $M_E$ and bound by the potential $V = \frac{k}{2}(z - z_{eq})^2$, where $z_{eq}$ - is its value at the equilibrium. In this approach the enzyme states are at thermal equilibrium, and so they have Boltzman spectra for $O_E$ energy at given temperature. Then the thermal fluctuations of $O_E$ energy result in $z$ value fluctuations, because of it the average tunneling coefficient will rise by many orders. For us the more interesting is another regime when the thermal fluctuations can be neglected, in comparison with the effect of $z$ variations induced by ordered excited state $G_j$ introduced above. Let's suppose for the start that this ordered state can be described as a shock wave (phonon) which has the constant amplitude $\Delta$ of barrier width oscillations during the excitation period $\tau$. Then the induced $z$ time dependence is:
$z(t) = z_{eq} + \Delta \cos \omega t$, as the result the tunnelling coefficient averaged over time is equal to:

$$T' = T(z_{eq}) \frac{1}{\tau} \int_0^\tau dt\, e^{-\frac{2\Delta \cos \omega t}{h}\sqrt{2m(U-E)}} \qquad (4)$$

where $\omega$ is $O_E$ frequency, $T(z_{eq})$ value is given by (1). $E$ - the kinetic energy of $m$ is taken to be the minimal possible for its motion, and $E=0$, $U \approx .3 \div .5$ eV can be taken arbitrarily.
In practical situations when $\omega^{-1} \ll \tau$, the integral is equal to $\tau I_o(\beta)$, where $I_0$ - modified Bessel function, $\beta$ is the numerical coefficient in the exponent rate of (4). Hence the average tunnelling coefficient is equal:

$$T' = T(z_{eq}) I_0\left(\frac{2\Delta\sqrt{2mU}}{h}\right) \qquad (5)$$

and doesn't depend on τ and ω values. The function $I_0(\beta)$ is approximated by the exponent with a reasonable accuracy, thereon $T'$ dependence on $\Delta$ is rather steep: for $\Delta=.2$ A the value of $T'$ changes by a factor 80, whereas the typical $z$ value is about 1A.

This calculations are applicable for the reactions which results only in a single H-bond formation, yet dNTP posses 2 or 3 H-bond vacancies which can be involved in biosynthesis simultaneously. Hence their confinement by the hot spot of dNTP-pol would be performed via the tunnelling of 2-3 protons via the different barriers. The analogous double proton tunnelling was already observed experimentally between organic molecules (Limbach et al., 2004); its computer simulation for G,C nucleotide dimer were published (Zoete, 2004). Eventually DNA fidelity $F$ will be proportional to $T'^{2\div3}$ in place of linear dependence on $T'$ for single p reactions; it means $F$ high sensitivity to the tunnelling parameters for dNTP reactions.

Now let's discuss the possible microscopic mechanism which permits for an enzyme to reduce the potential barriers between reactants. Until now in biochemical reactions this question is scarcely studied, but there are multiple indications that the electrostatic effects can be quite important for it (Benkovic, 2003). The experiments with the inorganic molecules show that the potential energy of H-bond would enlarge 5-6 times, if the analogous complex is anionic, the example is $(HF)_6$ versus $[F…H…F]^-$ (Joesten, 1974). The recent experiments and computer simulations for DNA nucleotides and their molecular analogs show the strong influence of electric charges configurations on proton transfer (Dabkowska,2004). The most important of them is the observation of barrier-free proton transfer (BFPT) in nucleotide dimers: if one nucleotide is anionic, then the rate of proton transition from other nucleotide or its analogue shows the practical absence of any potential barrier, which is of significant height and width for the same but neutral dimer (Gutowski,2002; Li 2001). The examples are anionic Uracyl, Glycine dymer; the other one most interesting for us is $C^-$,G dimer. In that case BFPT obtained for proton bound initially to N atom of G and attached finally to N atom of C anion (Li,2001). The calculated barrier height is about .05 eV, and this is about the value of proton kinetic energy, whereas for neutral G,C dimer this p-barrier is about .7 eV (Zoete,2004). Since the rate of proton transfer between nucleotides can be gained by the electrostatic effects, the analogous gain can be expected for H-bonds formation for the favourable configuration of electric field.

Basing on this premises we shall apply in our model such electrostatic mechanism of proton barrier reduction in DNA-pol hot spots. Let's consider first the case when only one H-bond should be formed between the hot spot and some molecule $A_P$, which for example is p-donor, and suppose that some negative charge $q$ is located near DNA-pol surface. Its electrostatic potential in lowest order approximation $V_E(\vec{r}) = -\dfrac{\varepsilon q}{r}$, where the coordinate $\vec{r} = 0$ in the charge centre; $V_E$ has the sign opposite to the potential $V_B$ of $A_P$ chemical bond p - $A_P$. The total potential for proton $U_T(\vec{r}) = V_B(\vec{r}) + V_E(\vec{r})$, and for large $q$ and small $r$ the potential $V_E$ can effectively reduce the barrier for p-transfer to the hot spot practically to zero level and possibly stimulate so H-bond formation. Namely, as was argued, the rate $R_H$ of H-bond formation is proportional to $T$ of (1), to estimate it in this case one can substitute $\sqrt{U-E}$ in (1) by the average $\overline{U}_T^{\frac{1}{2}}$ over barrier region which longitudinal size is *z*. Then as follows from (1) $R_H$ will grow exponentially with $\overline{V}_E$, as $\overline{U}_T$ falls down. If p-acceptor of hot spot is located near this charge (or it's simply the same ion), then the stable H-bond will be formed promptly between dNTP and DNA-pol. This charge can be static, yet in our model it's more plausible, if this uncompensated charge appears only during DNA-pol excitation period τ, and later is neutralized. In this case the neutral H-bond can be formed with high probability during this period, because of the reduction or disappearance of proton barrier. To bind dNTP with two or three H-bonds vacancies via such mechanism, the hot spot in excited state should include the complex array of positive and/or negative charges located near DNA-pol surface, so that all p-barriers can be reduced simultaneously. Below the simple variants of such configurations will be discussed.

In general the cited data evidence that experimentally observed gain of p-tunneling which BB and other models describe phenomenologically as the thermal fluctuations of barrier width, can have the electrostatic origin at microscopic level. Namely, both the dipole moments on enzyme surface, together with the thermal fluctuations of charge densities inside their volume, can produce the stochastic electric field, which due to large sizes of enzymes can be of significant strength; if such field has the suitable orientation near the barrier location it will suppress the initial p-transfer barrier (Hammes-Schiffer,2006). Naturally it doesn't exclude the alternative mechanisms of potential barrier reduction which act simultaneously with the electrostatic one, but here only this one will be regarded, because it seems most simple and appropriate for the regarded situation.

**C. Nucleotide Selection via Tunneling**

To find the optimal structure of hot spots $HS_j$ let's remind first the configuration of pDAs in dNTPs (fig .1). For any dNTP all its pDAs are located practically along a straight line with a distance about 1.8A between them; their exact values are given elsewhere (Watson et al., 1987). This oriented gap (vector) of the length about 3.6A is denoted **r**. Then their pDA spatial structure of nucleotides can be expressed with a good precision by the following symbols (Lowdin, 1963) :

C = {:H : :}; G = {: :H :H}; T={: :H (:)}; A={:H : (:H')}.

Here : and :H denote p-acceptor and H atom bound to p-donor correspondingly; the sign (:) denotes T lone pair - p-acceptor which stays vacant in stable A-T base pair of DNA (fig.2). (:H') denotes H atom of A which doesn't produce H-bond in DNA base pair and is separated from **r** for a distance about 1A. Yet as argued below, it probably can produce H-bond in the interactions with the hot spot of DNA-pol, so let's neglect for the moment this symmetry defect and take (:H')=:H. The configurations C and C′ ={: : :H} describe the same nucleotide rotated by the angle π , so that **r'**= **-r**; the same is true for G and G′={:H :H :}. The configurations {: : :} and {:H :H :H} aren't feasible for dNTPs, because of their too large electronegativity, or correspondingly positivity which can't be feasible for purine ring (Joesten , 1974).

Formally this pDA structure of nucleotides can be encoded by a three bit binary word $B_j = (b_1, b_2, b_3)$, where :H and : vacancies correspond to 1 / 0 values, for example C responds to $B_1 = (1,0,0)$. The complete $B_i$ set describes the natural numbers from 0 to 7, but $B_0 = (0,0,0)$ and $B_7 = (1,1,1)$ as was noted above, don't have feasible dNTP counterparts and should be rejected. As was also noticed, C and G are described by two $B$ elements, describing the same nucleotide rotated by π angle. These six elements $B_i$ constitute the symmetry group **B** of permutations which induces the dynamical symmetry of dNTP, DNA-pol interactions regarded below. Note that this group permits the existence of four fundamental nucleotides only; the meaning of this result will be discussed below.

To provide the optimal contact between dNTP and hot spot pDAs, the simplest $HS$ configuration of pDAs should be largely analogous to one of dNTP, so that its pDAs should be aligned along the straight line – oriented gap **s**, with one exception featured below. More precisely for given hot spot its pDA structure supposedly coincides mainly with pDA configuration of its base partner i.e. for $A^n$ its analogue of T, etc.. Its exact structure can be expressed by the addition (or subtraction) to it of few molecular fragments. In this vein we take that for $G^n$ exact pDA structure coincides with C one, and $C^n$ structure with G as well. To get the structure of $T^n$ hot spot, it should be added to A nucleotide configuration the p-donor complex located opposite of this O atom which stays vacant inside DNA (fig.1). Consequently T can form 3 H-bonds with $T^n$ geometric configuration $HS_2$.

Now let's regard in detail the mentioned defect :H' of A nucleotide (ddATP); this is H-C molecular fragment of purine (fig.1). Normally H-C fragment cannot form an H-bond, but this can occur if the carbon atom is connected with electronegative complex, the example is H-C≡N. In our case of ddATP it is connected with two N atoms and so in this configuration the creation of H-bond seems feasible. In DNA molecule this C-H fragment doesn't create H-bond with T base, first of all due to too large distance to p-acceptor of ddTTP (fig.2), but it can become possible for the suitable pDA structure of $A^n$ hot spot. The main structure of $A^n$ pDAs presumably coincides with T one, hence if the additional p-acceptor of HS $A^n$ situated at the necessary distance and angle from this pDAs, thereon ddATP will be also connected by 3 H-bonds with DNA-pol. Consequently pDA configuration of hot spots on DNA-pol surface can be expressed by the following symbols:

$C^n$= {: :H :H};  $G^n$= {:H : :};   $T^n$ = {:H : :H};   $A^n$={: :H :}.

which is the replica of corresponding C,G,T,A structure given above with the substitutions :H ↔ : and : ↔ :H. Hence this set is also isomorphic to the permutation group **B** which describes dNTPs symmetry. For such structure of DNA-pol hot spots all correct pairs of dNTPs and hot spots, like C,$C^n$, etc., are connected by three H-bonds, whereas all wrong pairs like A,$C^n$,etc., can produce two H-bonds maxima. This set of correct pairs is in fact the dynamical realization of dNTPs symmetry expressed by the group of permutations **B** described above.

Our consideration of all hot spots indicates that the geometrical structure of pDAs reproduces mainly the one of dNTP, i.e. pDAs are supposedly aligned along oriented gap **s** which supposedly lay on DNA-pol surface and separated by a distance about 1.8 A between pDAs. The only exception is $A^n$ for which one p-acceptor is disposed at the distance about 3.5A from p-donor and doesn't lay on DNA-pol surface along **s** and its extrapolation, but located about 1.5A higher. The optimal distance between pDA pairs of dNTP and hot spot for H-bonds formation is about 3A, hence their vectors **r** and **s** should be nearly parallel or antiparallel and their edges constitute the parallelepiped with the sides 3 and 3.6 A and orthogonal to DNA-pol surface. Hence the resulting structure of their H-bonds vacancies repeats the same structure of conjugated dNTPs shown on fig. 1. For example, their relative distances and orientations for $T^n$ are the same as for ddTTP H-bonds vacancies, but pDAs of $T^n$ are located on the opposed ends of dotted lines.

The proposed similarity of pDA structure for nucleotides and hot spots permits to assume that the proton barriers between them are analogous to such barriers in nucleotide dimers, i.e. are relatively high – about .5 eV. As we assumed above, such p-barriers can be suppressed electrostatically by the electric charges located in the hot spots near DNA-pol surface; this charges are induced in the hot spot $HS_j$ by DNA-pol excited state $G_j$. Because of large size and complicated structure of DNA-pol such fast rearrangement of charge positions seems quite possible. The data cited above, show that to remove such p-barrier completely it's enough to dispose the charge $e^-$ at the distance about $2 \div 3$ A from the location of nucleotide p-donor which isn't difficult to realize; the analogous features can be expected for nucleotide p-acceptor. This considerations permit us to propose the simple scheme of the barrier reduction: the excited state $G_j$ induces in the hot spot $HS_j$ the array of positive and negative charges which positions coincide with pDAs ones and charges signs $q^\pm$ are favourable for the electrostatic p-barrier suppression in given pDA. For example, for $T^n$ it can be the array $W_T = \{q_1^-, q_2^+, q_3^-\}$ with their positions coinciding with the positions of pDAs of $T^n$; $|q^\pm_i|$ absolute values are supposedly of the order of $e$. The electric fields of positive and negative charges can compensate each other at large distances, but for our pDA configurations this effect seems to be small. In fact such metastable dNTP – DNA-pol states can be bound not exactly by H-bonds, but in the presence of the electric charges this bonds can be the analogue of ionic bonds which is less sensitive to the system parameters. This question deserves further investigation, here we shall regard them formally only as H-bonds.

The regarded charges configurations are quite intricated, so it's worth to describe more simple but possibly less efficient ones. In this case each hot spot contains the charge of only one sign, which density $\rho_q$ is distributed optimally along all hot spot surface. Any regarded above charges configurations are asymmetric relative to the charge signs, because of the structures of their H-bonds, as $W_T$ example shows. In the regarded case the sign of hot spot charge is also selected in accordance with it. For example, $T^n$ includes two p-acceptors and one p-donor, so the charge of hot spot in this case is negative and so on. The data for double proton transfer in nucleotides show that such charge configuration also can enforce the formation of both H-bond types (Zoete, 2004).

In principle the regarded pattern of dNTPs interactions with DNA-pol via 3 H-bonds affords many different realisations of dNTP selection, here only the simplest one will be analyzed. The experimental data evidence that dNTP connected by 3 H-bonds inside DNA is bound more stronger than for 2 H-bonds, and possess larger stability against the thermal fluctuations (Watson et al., 1987). Let's suppose that in our case the total binding energy between dNTP and DNA-pol is proportional to the number of H-bonds, i.e. $E_B(n)=n E_H$, and take this energy of single H-bond equal to .17 eV, the typical H-bond energy in DNA. Let's compare the lifetimes $\tau_{2,3}$ of bound dNTP state for 2 and 3 H-bonds. Obviously the lifetime of any bound state is inverse proportional to the density of fluctuations with the energy higher than the binding energy and this density defined by Boltzman distribution. Hence the simple calculations give:

$$\alpha_{2/3} = \frac{\tau_2}{\tau_3} = e^{-\frac{E_B(3)-E_B(2)}{kT}} = e^{-\frac{E_H}{kT}}$$

which for room temperatures is about $10^{-3}$, i.e. the configurations with two H-bonds are essentially more unstable and short-living. Consequently, for the suitable cross-sections of dNTP thermal excitations, practically only dNTPs bound to DNA-pol by 3 H-bonds will be captured long enough to be transferred to the base template. The computer simulations of DNA-pol-β domains motion during 'induced-fit' conformational change don't contradict to such selection mechanism (Arora, 2004).

In terms of Computations Theory the formation or absence of each particular H-bond between dNTP and hot spot can be regarded as the classical query $Q_i$ of one item with two possible outcomes: *yes/*no ; the total number of items in the data base in this case is 3. The selection criteria adopted in our model – the creation of 3 H-bonds out of 3 possible vacancies – i.e. the triple coincidence; it is formulated as the algorithm of computation:

$S_Q = Q_1.and.Q_2.and.Q_3$

where $S_Q$ can have the values .true. or .false. (i.e. 1/0). In our formalism the structure of H-bonds vacancies for incoming dNTP can be expressed as 3-bit number $C_B=\{c_1,c_2,c_3\}$. If in a given replication step dNTP $D_j$, described by the element $B_j$, should be captured, then each of the three queries checks one $C_B$ bit, whether $c_i=b_i^j$ of $B_j$. Note that for the regarded dynamics all three queries are performed independently and simultaneously, hence this algorithm is the analogue of parallel computations. It's easy to check that this algorithm permits to identify only four different configurations of pDAs responding to four nucleotides. This result is strictly related with the fixed number of H-bonds, namely three, which in our model connects dNTP and its hot spot partner in case of correct identification. Its symmetry is described by the permutation group **B** introduced above; its elements correspond to four different nucleotides only. If to assume that the number of nucleobases can be less than 4, such hypothetic set of two or three nucleobases is obviously too small to describe the necessary diversity of aminoacids, whereas 5 or more of nucleotide species cannot be identified by means of 3 H-bonds only; it would need at least 4 H-bonds for their sorting out by DNA-pol. But in its turn it's possible only for the large size of molecules, which should carry this pDAs , i.e. larger than the pyrine ring of 6 atoms. But it will make such dNTP molecules inevitably more complicated and less stable. Therefore only the set of four nucleotides with particular pDA structure permits their unambiguous and simple identification and simultaneously possess the sufficiently simple molecular structure. We see that the optimal parameter of data base search – i.e.4 items in a single query, which were obtained in quantum search formalism (Patel,2001), are achieved also for our model system which isn't in pure quantum state. The complexity of system dynamics doesn't permit to prove that it isn't just a chance coincidence. However it's worth to notice that in this essence DNA-pol +dNTP is also the dynamic system with continuous spectra. In principle it also can have the resonance properties analogous to obtained for the system of coupled oscillators, and one can't exclude beforehand that just such properties permit to perform the optimal search of molecular data base for specific parameters of our model (Grover, 2002).

## 3.  Concluding Remarks

We regarded the simplest mechanism of dNTP identification by DNA-pol, but more complicated and efficient mechanisms can exist which will be considered in forthcoming paper. In our model it supposed that 2 or 3 protons transfer between DNTP and DNA-pol occurs independently of each other, however there are some indications of dynamical correlations between this protons (Zoete,2004).  There is also the tempting possibility that all 'hot spots' are identical, and each $G_j$ state tunes them differently for the binding of particular dNTP sort, but it demands quite complicated mechanism. We didn't study here the suppression of nucleotide background produced by other molecules which can also form H-bonds, first of all $H_2O$, the same is true for other dense media effects. It is quite an important problem which demands a detailed study, primarily to understand whether the geometry of DNA-pol allows to protect the vicinity of hot spots from direct contacts with water molecules. As was noticed in the introduction, to make DNA-pol performance truly efficient it's enough for the regarded preselection to have fidelity of the order $10^{-2 \div 3}$, so the additional selection can be fulfilled after the 'palm' closes and DNA base pair biosynthesis starts. The regarded mechanism of preselection doesn't exclude the subsequent action of other selection mechanisms which start to act after the palm of DNA-pol closes (Kunkel, 2004 ).

In this paper we don't discuss how DNA-pol identifies initial DNA base template for which the nucleotide partner should be selected, accepting it at this stage 'ad hoc'. Naturally this question deserves the serious investigation, from our results it seems reasonable to assume that such mechanism can be similar to one studied here, i.e. that DNA-pol separates the nucleobase templates also via the different structures of their H-bonds vacancies. Note however that in this case their identification should be much simpler, because the position of template nucleobase relative to DNA-pol is practically fixed.

If further investigations will support the described model of the information transfer in biomolecules, it can help also to construct the molecular computers which will be more efficient for calculations of protein properties than the standard computers. Here the simple example of such computations, performed during the nucleotides identification, was regarded. In general the regulation of proton tunnelling in chemical reactions opens the new perspectives in bioelectronics and biocomputing. This process is the analogue of silicon tunnel transistor, in which the control voltage determines the height of potential barrier for electrons and hence the output current. This effect was applied for the realization of computer 'control gate'; in biocomputing it can be realized possibly by a variety of enzymatic reactions. Remind that the tunnelling in semi- and superconductors is extremely sensitive to the state of its environment, such sensitivity permitted to construct the large family of the perfect sensors devices. Analogously enzyme tunnelling of protons can be sensitive not only to the temperature as was already demonstrated, but to other parameters, such as the external pressure. The clear indication of its influence on the rate of some chemical reactions via the proton tunnelling mechanism are obtained already (Northrop, 2002). By the analogy it seems worth to study whether the proton tunneling in biochemical reactions can depend on other external parameters like the magnetic field, electromagnetic waves frequency and intensity, etc.. In principle such effects can open the new prospects in the external control and the information transfer in Bioelectronics.

Hammes-Schiffer S. (2006) Hydrogen Tunneling and Protein Motion in Enzyme Reactions Acc. Chem. Res. 39: 93-100

Joesten M., Shaad L. (1974) Hydrogen Bonding. New York: Dekker.

Kohen A., Klinman J.P. (1999) Hydrogen tunneling in biology. Chem.Biol. 6:R191-R198.

KortemmT., Baker D. (2002) Simple Physical model for binding energy hot spots in protein complexes. Proc.Natl. Acad. Sci.USA 99 (22) :14116-14122

Kunkel T. (2004) DNA replication fidelity. J. Biol. Chem. 279:16895

Li X., Cai Z., Sevilla M. (2001) Investigation of Proton Transfer within DNA Base Pair Anion and Cation Radicals. J. Phys. Chem. B 105: 10115-10123

Limbach H.H., Klein O., Del Amo J.M.L., Elguero J. (2004) Kinetic Hydrogen/Deuterium Isotope Effects in Multiple Proton Transfer Reactions Z. Phys. Chem, 218: 17-49

Lloyd S. (2000) Quantum search without entanglement. Phys. Rev A61: 010301

Lodish H., Berk A., Zipursky S.L., Matsudaira P. (2000) Molecular Cell Biology. 4th. Ed. New York: Freeman.

Lowdin P. (1963) Proton Tunneling In DNA. Rev. Mod. Phys. 35, 724

Masgrau L, Roujeinikova A, Johannissen LO, Hothi P, Basran J, Ranaghan KE, Mulholland AJ, Sutcliffe MJ, Scrutton NS, Leys D. (2006) Atomic description of an enzyme reaction dominated by proton tunneling. Science 312:237-241

Northrop D.B. (2002) Effects of high pressure on enzymatic activity. Biochim. Biophys. Acta. 2002 1595:71-79

Patel A. (2001) Quantum algorithms and the genetic code. Pramana 56: 367 ; Quant-ph/0002032

Sawaya M.R. et al. (1997) Crystal structure of human DNA polimerase β complexed with gapped DNA : evidence for Induced Fit mechanism. Biochemistry 36 :11205-11215

Watson, J.D., Hopkins, N.H., Roberts, J. W., Steitz, J.A. and Weiner A.M. (1987). Molecular Biology of The Gene. The Benjamin/Cummings Publishing. Co. California/USA.

Zoete V., Muwly M. (2004) Double Proton Transfer in Isolated and DNA-embedded Guanine – Citosyne base pair. J. Chem. Phys. 121: 4377-4388
## Figure Captions

Fig.1 The structure of p-donors and p-acceptors for free nucleotides A - Adenyne, T - Thymine,G - Guanine, C - Cytosine;
H-bonds vacancies denoted by green dotted lines

Fig.2 The structure of H-bonds in nucleotides bounded in DNA,
H-bonds between nucleotides A-T, C-G denoted by gray dotted lines

Fig. 1

Paired nucleotides

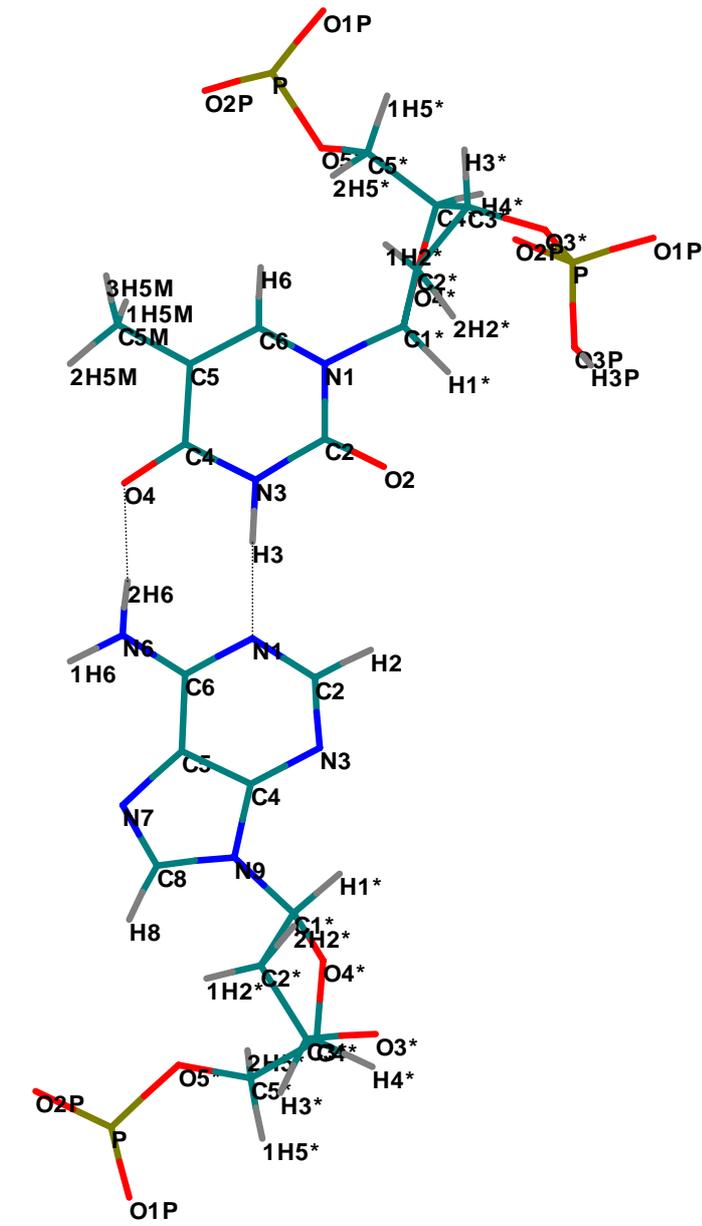 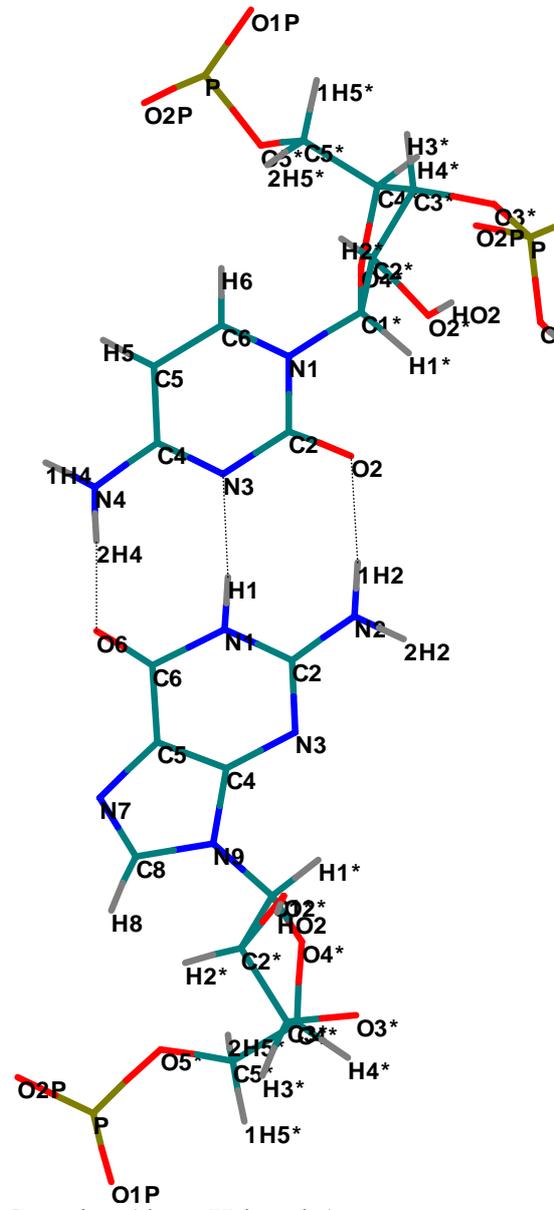

Adenine : Thymine (two H - bonds)    Guanine : Cytosine (three H-bonds )

Fig. 2